\documentclass[11pt]{article}
\usepackage[a4paper, total={7in, 10in}]{geometry}

\usepackage{graphicx}
\usepackage{helvet}
\usepackage{authblk}
\usepackage{xcolor}
\usepackage[colorlinks=true,linkcolor=blue,citecolor=blue,urlcolor=blue,allbordercolors=white]{hyperref} 
\usepackage{amsmath} 
\usepackage{amssymb} 
\usepackage[super,comma,sort&compress]{natbib}
\makeatletter
\renewcommand{\maketitle}{\bgroup\setlength{\parindent}{0pt}
\begin{flushleft}
  \textbf{\@title}
  
  \@author
\end{flushleft}\egroup}
\makeatother

\title{Mapping gene expression dynamics to developmental phenotypes with information entropy analysis}
\date{}

\author[1,$\dagger$]{Ben Ansbacher}
\author[1,2,$\dagger$]{Malachy Guzman}
\author[2,*]{Jordi Garcia-Ojalvo}
\author[1,*]{Arjendu K Pattanayak}

\affil[1]{Department of Physics and Astronomy, Carleton College, Northfield, MN, USA}
\affil[2]{Department of Medicine and Life Sciences, Universitat Pompeu Fabra, Barcelona, Spain}
\affil[$\dagger$]{These authors contributed equally to this work.}

\begin{document}

\maketitle

\section*{Abstract}

The development of multicellular organisms entails a deep connection between time-dependent biochemical processes taking place at the subcellular level, and the resulting macroscopic phenotypes that arise in populations of up to trillions of cells. A statistical mechanics of developmental processes would help to understand how microscopic genotypes map onto macroscopic phenotypes, a general goal across biology. Here we follow this approach, hypothesizing that development should be understood as a thermodynamic transition between non-equilibrium states. We test this hypothesis in the context of the fruit fly, \textit{Drosophila melanogaster}, a model  organism used widely in genetics and developmental biology for over a century. Applying a variety of information-theoretic measures to public transcriptomics datasets of whole fly embryos during development, we show that the global temporal dynamics of gene expression can be understood as a process that probabilistically guides embryonic dynamics across macroscopic phenotypic stages. In particular, we demonstrate signatures of irreversibility in the information complexity of transcriptomic dynamics, as measured mainly by the permutation entropy of indexed ensembles (PI entropy). Our results show that the dynamics of PI entropy correlate strongly with developmental stages. Overall, this is a test case in applying information complexity analysis to relate the statistical mechanics of biomarkers to macroscopic developmental dynamics.
  
\section*{Significance statement}

Using arguments drawn from non-equilibrium thermodynamics, we hypothesize a correlation between gene expression dynamics during development and corresponding macroscropic/phenotypic descriptions of the developing organism. We test this hypothesis in the context of the fruit fly, \textit{Drosophila melanogaster}, using a variety of information-theoretic measures to analyze public transcriptomics datasets of whole fly embryos during development. We find that gene expression dynamics can be understood as a process that probabilistically guides embryonic dynamics across macroscopic phenotypic stages. In particular, we observe signatures of irreversibility in the information complexity of transcriptomic dynamics, as measured mainly by the permutation entropy of indexed ensembles (PI entropy), as well as a strong correlation with a measure of developmental stage.

\section{Introduction}
        
The development of a living system from a single cell (the zygote) to a fully formed organism can be interpreted as the mapping over time of a large number of microscopic degrees of freedom at the molecular level, the genotype, into a smaller number of macroscopic features at the supracellular level that determine the organism's form and function, the phenotype.
This process is driven by carefully choreographed gene expression in space and time, the details of which have been mostly unraveled by developmental and molecular biologists over the last few decades \cite{wolpert_principles_2019}.
It is now known, for instance, that cell-fate decisions and morphogenesis are two tightly coordinated processes at different scales (cells and tissues, respectively), whose outcome is highly predictable at the scale of the organism \cite{keller_developmental_2002}.
The specific regulatory mechanisms underlying this coordination are, however, highly contingent on both the developmental stage of the organism, and the species to which the organism belongs.
The question still remains of whether there are fundamental principles underlying this developmental genotype-phenotype mapping \cite{manrubia_genotypes_2021}.



For instance, it remains puzzling how the remarkably consistent organization exhibited by developing organisms emerges from random sources of energy, as well as via biochemical processes that are themselves stochastic \cite{toppen_noise-driven_2025}. We argue that one way of attacking this puzzle is through a thermodynamic and statistical physics framework, using concepts from information theory and complexity theory. Specifically, we hypothesize that development proceeds as a time-directed process affecting both the organism and its environment simultaneously, driven by transitions—sometimes smooth and occasionally abrupt—between different kinds of non-equilibrium thermodynamic states.  Such a thermodynamic trajectory should map to a developmental trajectory and be independently visible in (a) microscopic system state properties, 
(b) macroscopic or coarse-grained system properties, and 
(c) the thermodynamic exchange with the environment. 
While the question of experimentally measuring (c) is under active investigation \cite{ghosh_developmental_2023}, here we present a test case connecting (a) and (b) using publicly available bulk transcriptomics (RNAseq) data from whole \textit{Drosophila melanogaster} embryos, measured at multiple stages of development \cite{hooper_identification_2007,arbeitman_gene_2002}.
We use this gene expression data as the microscopic observable, to which we apply our information-theoretic measures. We then compare these quantifiers with a qualitative measure of developmental stage as the macroscopic observable. 

Our results show signatures of irreversibility over the entire period of embryonic development, visible in information-theoretic representations of the gene expression dynamics, which we sketch here and discuss more in detail below.
In particular, the PI entropy \cite{aragoneses_permutation_2023}, or $\Pi(t)$, works by creating a distribution of genes indexed by their rank-order in expression value at the final timepoint. This is the designated reference, or `equilibrium,' distribution. At every other timepoint, each gene's expression value is moved up or down relative to this indexing, resulting in a new distribution with fluctuations relative to the reference one.
As described in the sections that follow, $\Pi(t)$ quantifies the behavior of these fluctuations as a function of time, displaying thermalization behavior.
Specifically, we find that this distribution is initially well-shuffled and wildly fluctuating relative to the final distribution, and settles through continued shuffling as time progresses, resulting in increasing smoothness relative to the final time point.
In other words, the ensemble relaxes as the shuffling happens on progressively smaller scales along both the expression and indexing axis as the system approaches its `final' state at the end of development.  

We also find that the information-theoretic dynamics described above correlate strongly with a phenotypic measure of the macroscopic developmental stage. The correlation is remarkable given the relative crudeness of the data and the degree to which we are using proxies across multiple scales of coarse-graining for both microscopic and macroscopic properties. The dynamics of macroscopic and microscopic metrics also exhibit the characteristic features of a thermodynamic transition between two different nonequilibrium states. 

Taken together, these analyses of gene expression dynamics unveil statistical mechanical trajectories that both show irreversible convergence of the ensemble dynamics of the microscopic determinants of development and correlate with macroscopic properties.
The time evolution of these two properties is consistent with the broader idea of development as a sequence of transitions between non-equilibrium states. In particular, it seems appropriate to consider expression levels to be playing the role of `energy', while the PI plays the role of the related `entropy'. 

We also use other ways of filtering the data to probe details of the thermodynamic trajectory.
This includes analyzing subsets of genes that display high dynamical variability, using eight groups of tightly co-expressed/co-regulated gene that have been previously identified \cite{hooper_identification_2007}.
These gene subsets show a variety of expression dynamics within the overall behavior described above: some start high in expression and decrease steadily as time progresses, others start and end at low expression values with a blip of activity in between, and some others start low and end high.
We see in these individual subset dynamics, as in the global dynamics, that $\Pi$ tends to be higher when mean expression values are higher. The results obtained from the analysis of both the global transcriptome and the expression of the gene subsets described above are consistent with more `entropy' being available at higher `energy' levels.

The overall irreversibility in the transition between the initial and final developmental states thus proceeds via a settling to lower mean expression ('dissipation to lower energy') in gene expression space, and a separate `thermalization' time-scale  visible in $\Pi$ dynamics, where the ensemble expression dynamics is seen to settle down as it approaches the final time, without any overall change in mean expression.

We start in Methods with the theoretical framework, including the connection between information theory and thermodynamics, and spelling out in more detail our metrics and their interpretation, along with the characteristics of the gene expression data itself. In Results, we present global irreversibility and correlations with phenotype behavior, along with a short discussion on what we learn from a focused look at temporally filtered `active' gene subclasses, and what this means for underlying expression dynamics in gene regulatory networks.
We also discuss what we learn by filtering by gene identity, and what the overall dynamics say about developmental stages and irreversibility. We conclude with a short summarizing discussion of the interpretations of these results and on prospective work as well as the broader applicability of information-theoretic techniques in developmental biology.

\section{Methods}
\label{sec:methods}

\subsection{Statistical mechanical approach and irreversibility}
While physical entropy production increases with time in a developing organism, biological organization also increases. There is no inherent contradiction, since any particular organism trades its increasing internal order and associated range of coherent use of energy in time and space for even greater disorder produced or exchanged with its environment \cite{schrodinger_life_1944}. Considering this exchange between the system and the environment means that within a thermodynamic framework, development can be understood as a thermodynamic trajectory between states with different energetic and entropic properties. That is, the biological system undergoes an irreversible evolution with transitions 
between different non-equilibrium states. Thus, changes in energy usage and entropy production properties should correlate with biological changes, visible in developmental stages.  

 In order to construct the statistical mechanics of this developmental and thermodynamic trajectory, we have to consider two sensibly well-separated levels of description of the biological system. Micro- and macro-states can be descriptions of the organism \cite{garcia-ojalvo_towards_2012} ranging from single-cell expression data all the way to organism-level phenotypes.
 The degree of `coarse-graining' in our description of the system determines the microscopic variables, and hence our statistical `microstates', in turn delimiting our biological or thermodynamic macrostates.
 It is intuitive that phenotypic descriptions should correspond to biological macrostates.
 We note that a separate quantitative description exists in the genuinely thermodynamic properties such as the energy and entropy exchange of the developing organism(s) with the environment.
 This is not, however, the approach that we follow in this paper.

Development should be visible in the dynamics of (a) microscopic system state properties evolving or available to us as $\rho(\mu(t))$, where $\rho$ is an ensemble or distribution over some generically labeled observable $\mu(t)$.
Further, this dynamics can be quantified by (b) macroscopic or coarse-grained system properties generically labeled as $M(t)$, as well as (c) thermodynamic exchange with the environment, generically $E(t)$, where these are properties can be measured using tools like calorimeters and spectrometers.
A statistical mechanics of biological development should in principle relate $\mu(t)$ and $M(t)$, the thermodynamics of development should relate $M(t)$ and $E(t)$, and our hypothesis should connect all three.
Detailing the connections between these levels of description is in general an arduous task.
However, we can make some progress by taking advantage of the fact that development is a fundamentally irreversible process between phenotypic stages, and hence that all three levels of description -- statistical properties of $\mu(t)$, as well as appropriate $Mt)$ and $E(t)$ directly -- should show this irreversibility. Thus we start by searching for appropriate descriptions at relevant scales that indeed show this property. Further progress would then arise from confirming that such behavior is indeed visible across scales of description, as well as exploring how the correlations between the micro- and macroscale descriptions change as a function of the environment.

In particular, we have seen in physics \cite{latora_kolmogorov-sinai_1999,falcioni_production_2005,aragoneses_permutation_2023} that if we use appropriate information-theoretic characterizations $I[\rho(\mu)]$, then the dynamics $I(t)$ correlates with the dynamics of the statistical entropy of the system in physical ensembles $\rho$, and by extrapolation with the system's thermodynamic entropy. We adapt this to the context of developmental biology to hypothesize a correlation between information-theoretic dynamics $I[\rho(\mu(t))]$ and a measure $M(t)$ of macroscopic development as a function of time. $M(t)$ should itself correlate with energy, entropy production, and material exchange $E(t)$ with the environment.  
Our initial task reduces to finding an observable $M(t)$ that is a good proxy for the macroscopic (phenotypic) irreversible dynamics, as well as to identify a property $I[\rho(\mu(t))]$ of the microscopic data that correlates with the macroscopic data, and in particular displays irreversibility itself.

First, we must clarify what we mean by `irreversibility' in this context. Consider the dynamics or arrow of time defined by the relaxation of any non-equilibrium initial ensemble to equilibrium in physics, for example a sharply initialized Liouville ensemble in classical physics. More detailed references and foundational discussion of the behavior of the entropy in this context can be found in Refs.~\cite{latora_kolmogorov-sinai_1999,falcioni_production_2005,aragoneses_permutation_2023}, and we summarize the intuitive picture here. We expect that as such an ensemble `thermalizes' (goes to its equilibrium steady state in phase space), it does so with an overall increasing entropy and a distinct sense of direction in time. That is, when a film of the ensemble dynamics is run backwards or forwards in time, we can visually intuit the difference through our expectation that in forward time, the ensemble gets increasingly disordered and spread through phase space. Quantitatively, we can see the difference in the behavior of macroscopic properties such as the entropy, measured both as a statistical property of the ensemble as well as `thermodynamic' changes that would be visible in a laboratory. This clear distinction between the forward and backward evolution of an ensemble is the essence of irreversibility. Since the formal calculation of the entropy of a Liouville ensemble is extremely difficult, a good measure of irreversibility is a proxy measure that collapses the full ensemble dynamics into a scalar function of time that either decreases or increases with time overall (not necessarily monotonically). We also seek measures with dynamics that correlate with patterns or time-scales for microscopic trajectories, the macroscopic ensemble behavior, or both \cite{aragoneses_permutation_2023}. An information-theoretic method to capture this in the Liouville ensemble quantifies the growth of differences between initial conditions or the resulting loss of memory of initial correlations between ensemble members.

Developmental biology also has the concept of an arrow of time, but fundamentally inverted in perspective. That is, the striking difference between a developmental biologist's and a physicist's intuition about irreversibility is that a physicist's arrow of time and growth of disorder is visible in how a structured egg gets scrambled, whereas the biologist's arrow is visible in how a relatively featureless egg develops into a highly structured organism. While both perspectives distinguish the future from the past, the physicist's perspective measures loss of correlations and irreversible transitions away from an initial state, while the biologist's perspective sees a growth of correlations and irreversible change of state relative towards a final well-defined state \cite{kondev_biological_2025}. 

We thus seek as a signature of irreversibility an information-theoretic property of the microscopic biodynamics that relaxes or thermalizes as a function of time, as in physics. during Development however, instead of the reference state being the initial state, the final state should be the appropriate reference state against which dynamics are measured \cite{kondev_biological_2025,gerhart_cells_1997,koslover_many_2025}. Therefore, we seek to take biodynamical trajectory ensembles in microscopic space, `collapse' or project these dynamics onto the behavior of an `entropy-like' measure, and examine the dynamics of this measure, in particular to see if these indicate a definite transition between the initial and final states with an overall change in the ensemble's dynamics akin to `thermalization' in physics. 

More technically, a good measure for a given system or set of trajectories should allow us the maximum discrimination between different microstates when mapped to a single macroscopic property, i.e., it should vary over as wide a range as possible (and as smoothly as possible as a function of microstate and time) to be an appropriate candidate to correlate with a physical dynamical observable for the ensemble. Once irreversibility within the dynamics of such a measure is found, we can hope to correlate this with macroscopic measures of irreversibility that are more easily measured or approximated, but still have to be chosen with care.

In this paper, we work with a microscopic distribution $\rho(\mu(t))$ constructed using publicly available bulk RNAseq data, which we characterize using the mean expression and particular information-theoretic constructions: The Shannon entropy, a specialized form of the Kullback-Leibler (KL) entropy, as well as the PI entropy discussed below in more detail. We also use the ensemble averaged developmental stage (discussed in more detail below) from the same experiments \cite{hooper_identification_2007,arbeitman_gene_2002} as our macroscopic phenotypic property $M(t)$. The question of measurement and characterization of $E(t)$ in the laboratory remains to be understood and explored in future work.

\subsection{Data acquisition and processing}\label{data}

We analyze public bulk RNAseq data from \textit{Drosophila} embryos spanning 24 hours of development.
Specifically, we use the Whole Genome Drosophila Embryogenesis Time Course dataset available from the NCBI GEO database \cite{barrett_ncbi_2012}, under accession number GSE6186.
This dataset contains the bulk transcriptome of \textit{Drosophila melanogaster}, tracking the expression levels of 11,456 genes at 30 time points covering the entire 24-hour period when the embryos develop into larvae \cite{arbeitman_gene_2002}. Independent embryos were harvested for data collection at each time point. The proportion of embryos in a particular embryonic stage (as reported by the dataset) was measured alongside relative gene expression values at each time point. For the first 6.5 hours of measurements, there are overlapping time points, but all measurements after hour 6.5 are taken at 1-hour intervals. For example, there is data taken between hour 1 and 2, and as well between hours 1.5 and 2.5; we call the latter `half time' points. Though care was taken to minimize overlap between measurements, in the bulk of our analysis, we remove these 'half time' data points and note specifically when they are included. 

The gene expression values in the dataset were originally reported as $\log_2$-transformed expression ratios (relative expressions) with respect to a common reference sample, in order to make expression levels cross-comparable. In some of our computations, we reverse the $\log_2$ transform to give a larger dynamic range and ensure the approach KL metric ($A_{KL}$) is well-defined. More detailed descriptions of data collection techniques and procedures can be seen in \cite{hooper_identification_2007} and \cite{arbeitman_gene_2002}. 

Since the data do not track the gene dynamics of a collection of the \emph{same} embryos over time, it is not exactly analogous to trajectories in physics. Instead, it consists of averages over the collected embryos at each time point, normalized to a reference taken over the entire lifetime. In that sense, the data is more aptly characterized as being collected from an ensemble over multiple individual embryo trajectories, yielding the transcriptomic state trajectories averaged over individual genetic variation, throughout development. 

Data were analyzed using the statistical package R (version 4.2.0) and the python packages NumPy (version 1.26.4) and SciPy (version 13.1). Kernel density estimates were computed with the R package ggplot2 (version 3.5.1) and mutual information values were computed with the R package infotheo (version 1.2.0.1). In order to compute mutual information, we dicretized our data into $\lceil N^{1/3}\rceil$ bins, where $N$ is the number of observations for the given metric.


\subsection{Information complexity metrics}

\label{sec:metrics}
In what follows, we describe the various ensemble-averaged properties of the time-dependent gene expression vector $\mu(t)$ over the $N = 11,456$ genes, each labeled by the index $i$. The simplest quantifier that we define is the average $\sum_i \mu_i(t)/N$, which we term `mean expression level' $\bar\mu$. To quantify the complexity of the shape of the probability distribution of expression values $P(\mu)d\mu$, we measure the Shannon entropy $H =-\int P(\mu)\ln (P(\mu))d\mu$; for our discrete data this is calculated by defining a histogram of probability of expression levels $P_n$ over the $nth$ neighborhood of $\mu$ of width $\Delta \mu$, i.e., by binning along the expression axis in bins of size $\Delta \mu$ and instead computing the discrete Shannon entropy $H= -\sum_n P_n\ln P_n$.  

In order to quantify how the transcriptome approaches its final state, we apply the Kullback-Leibler (KL) approach distance \cite{beck_thermodynamics_1995}, defined as:
\begin{equation}
    A_{KL}(t_j) = -\ln \frac{ \left[\vec{\mu}(t_j) \cdot \vec{\mu}(t_f)\right]^2}{|\vec{\mu}(t_j)|^2 \cdot |\vec{\mu}(t_f)|^2},
    \label{eq:akl}
\end{equation}
which compares the time-dependent $N$-dimensional gene-expression vector $\vec{\mu}$ at a given time $t_j$, with its value at the final time point $t_f$, which we consider the final state of the system.
We use the dot product between the vectors as a measure of the distance between them. The normalization by the length of the vectors and the negative logarithm ensures that this metric is $0$ for a perfect overlap between the two gene-expression vectors, and increases monotonically as the dot-product decreases (i.e., as the correlation between the vectors decreases). We thus expect this measure to decrease over time as the transcriptome trends towards its final state.

Finally, we also use the Permutation entropy of an Indexed ensemble (PI or $\Pi(t)$). This recently proposed measure \cite{aragoneses_permutation_2023} tracks how smoothness in an ensemble evolves as a function of time, in the physics case typically going from an initially prepared smooth distribution to a complex one as a function of time. The method starts from an ensemble of trajectories that are initially indexed along a coordinate and uses symbolic or ordinal analysis \cite{bandt_permutation_2002,leyva_20_2022} to compute how this indexing becomes braided over time due to individual trajectory dynamics. The change in smoothness over the ensemble of trajectories as measured by $\Pi(t)$ was shown to behave as a good proxy for the entropy dynamics of a non-interacting ensemble of chaotic maps. The degree of braiding --- relative to a chosen and prepared `un-braided' state --- provides both a sense of distance and direction in ensemble space, and in particular a measure or growth or loss of correlations.
In using this technique, we treat the gene expressions data as an ensemble of trajectories, and choose the last time point as our reference point for `smoothness'. This amounts to designating the final state to be when the gene expression ensemble is most `at equilibrium', and computing the time development of an information-theoretic distance from this state, as measured by deviation from its referenced smoothness. The coordinate on which we sort the data---that is, choose neighborhoods in which to define smoothness---is the gene expression value $\mu_i(t_f)$ at the final time. 

Specifically, to compute the PI entropy here, we start by indexing all genes according to their expression values at the final time point ($t=t_f$). The gene with highest expression at $t_f$ is assigned position $i=0$, the next most-highly expressed is assigned $i=1$, and so on. We then consider the indexed ensemble $\tilde{\rho}$ for each time point $t \in \{t_0,t_f\}$, and at each $t$ compute the degree of `braiding' or shuffling by comparing triplets of expression values as we move along $i$ from $i=0$, defining ordinal patterns for each of these sequential triplets of gene-expression levels. For example, if $\mu_0<\mu_1<\mu_2$ we record $012$, if $\mu_1<\mu_0<\mu_2$ we record $102$, etc.
We then obtain at each time $t$ the relative probabilities of the seven possible patterns ($111,012,021,102,120,201,210$), and finally the corresponding PI entropy for the ensemble
\begin{equation}
\Pi = - \sum_i P_i \log P_i,
\end{equation}
where $i=1,..., 7$ denotes each of the seven patterns listed above. Work with physical dynamical maps \cite{aragoneses_permutation_2023} and flows \cite{prasad_pi_nodate} has shown that $\Pi(t)$ measures the change in complexity of an ensemble relative to the reference ensemble (in those cases taken to be the initial ensemble, in contrast with our case). In those analytically computed dynamics, the initial or reference smoothness thus yields a transition from an initial completely ordered distribution where $P_{012}=1$ and the rest of the ordinal patterns have zero probability, to more complicated distributions, as might be intuitive. 

There is, however, a critical difference in this application to real data, due both to noise as well as the fine-scaled nature of the differences between the expression values. In particular, here small differences in expression may not reflect actual and significant change, but rather amount to essentially unaltered expression levels \cite{sun_complexity_2010}. Inspired by this, we impose a `tie threshold' that renders a set of $3$ expression values as equal, denoted by the word $111$, if any one of their differences is less than or equal to a threshold $\sigma_{\text{tie}}$. A gene's expression level in a set is regarded as unchanged if the respective difference was below a specified threshold, $\sigma$, set to $\sigma = 0.26$. This  value corresponds to the determined mean absolute difference plus one standard deviation between expression values across all transcription vectors. 

Further, in considering $\Delta^i_\mu(t)  =\mu_{i}(t) - \mu_{i+1}(t)$ between consecutive expression values, recall that all expression levels lie between $\mu_{min} \leq \mu \leq \mu_{max}$ (in our case $\mu_{min} = 0.04$ and $\mu_{max} = 23.6$ with the $\log_2$ transformation undone), so that differences between expression levels scale as $(\mu_{max}-\mu_{min})/N$. Given our large $N$, the typical $\Delta^i_\mu(t)$ for any distribution ranked by expression levels is quite small. This, along with our use of a `tie' threshold, implies that our metrics work differently here than for the deterministic chaos cases previously studied \cite{aragoneses_permutation_2023}. There, the signature of smoothness is in the initial state, which is completely defined by the $P_{012}$ population, and the system deviates from this as a function of time. Here, the expression data is inherently  noisy enough that any sufficiently fine-grained or smooth expression curve when filtered through the tie threshold yields an ordinal pattern distribution dominated by the $P_{111}$ population rather than by $P_{012}$ when we sort/index by expression value as we do for the final time point. Given the large number of genes, the distribution is indeed quite fine-grained and smooth, whence $\Pi(t)$ in this system serves as a measure of a departure from this sort of smoothness relative to a reference time. 
This relationship between $\Pi$ and $P_{111}$ is visible in Fig.~\ref{fig:PI-P111}.

The behaviors of these separate measures add up to tell a more comprehensive story than any one measure by itself. The mean expression level $\bar\mu$ acts as an equivalent of an energy-like property in physics, which makes intuitive sense particularly as we consider the energy requirements to express a gene. The Shannon entropy $H$ in turn characterizes the complexity of the shape of the distribution of gene-expression levels when sorted by expression value, crudely measuring the degree to which the expression levels are spread out or localized across genes. This metric, however, does not pay attention to the identity of any gene at a given expression value, thus ignoring `exchanges' in expression levels, for example. The Kullback-Leibler distance metric $A_{KL}$ is easy to visualize as resulting from evolution in a high-dimensional vector space, but does not provide an intuition about direction or what a distance means, though it can identify change easily. Finally, the PI entropy $\Pi$ is subtler to understand but has the advantage of being able to associate a direction in time with the change, as we see below. 

We also note that our choice of coordinate for indexing is implicitly setting up gene-expression levels as a proxy of how energy might behave in a physical situation, thus implicitly the PI resulting from this indexing is a proxy related to entropy.
We now show how these quantities behave for the developmental data considered in this paper.

\section{Results}
\label{sec:results}

\subsection{Gene expression dynamics: Irreversibility and correlation with developmental stages}


\begin{figure}[h]
    \centering
    \includegraphics[width=0.9\linewidth]{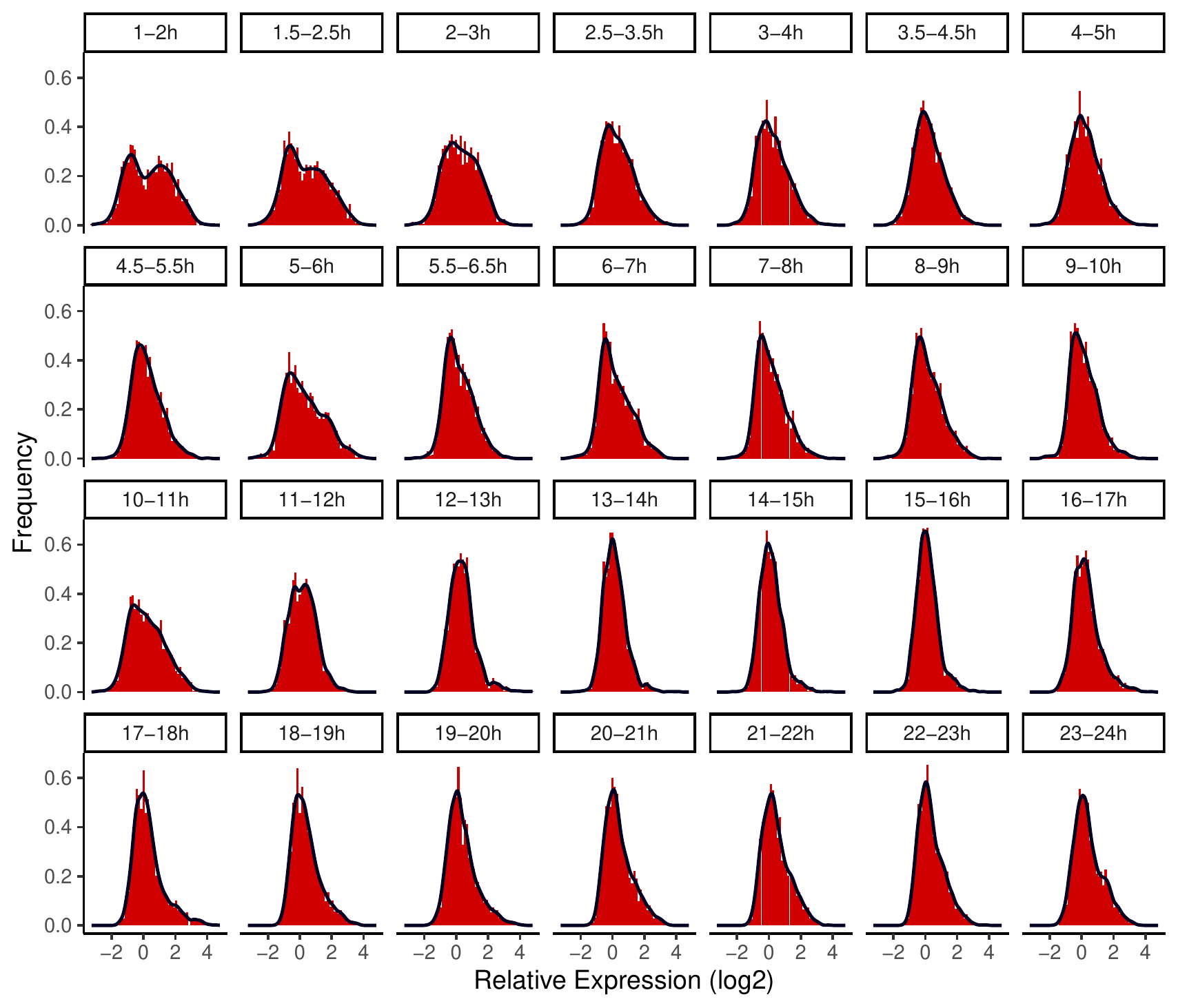}
    \caption{Time evolution of the expression distribution. These panels show the histograms of gene expression as a function of expression level as a function of timepoint. The data includes genes both with and without highly variable expression. The black curves (`envelopes') are obtained from standard kernel density estimates (see Methods).}
    \label{fig:DistributionsByTime}
\end{figure}
\begin{figure}[htbp]
    \centering
    \includegraphics[width=0.8\linewidth]{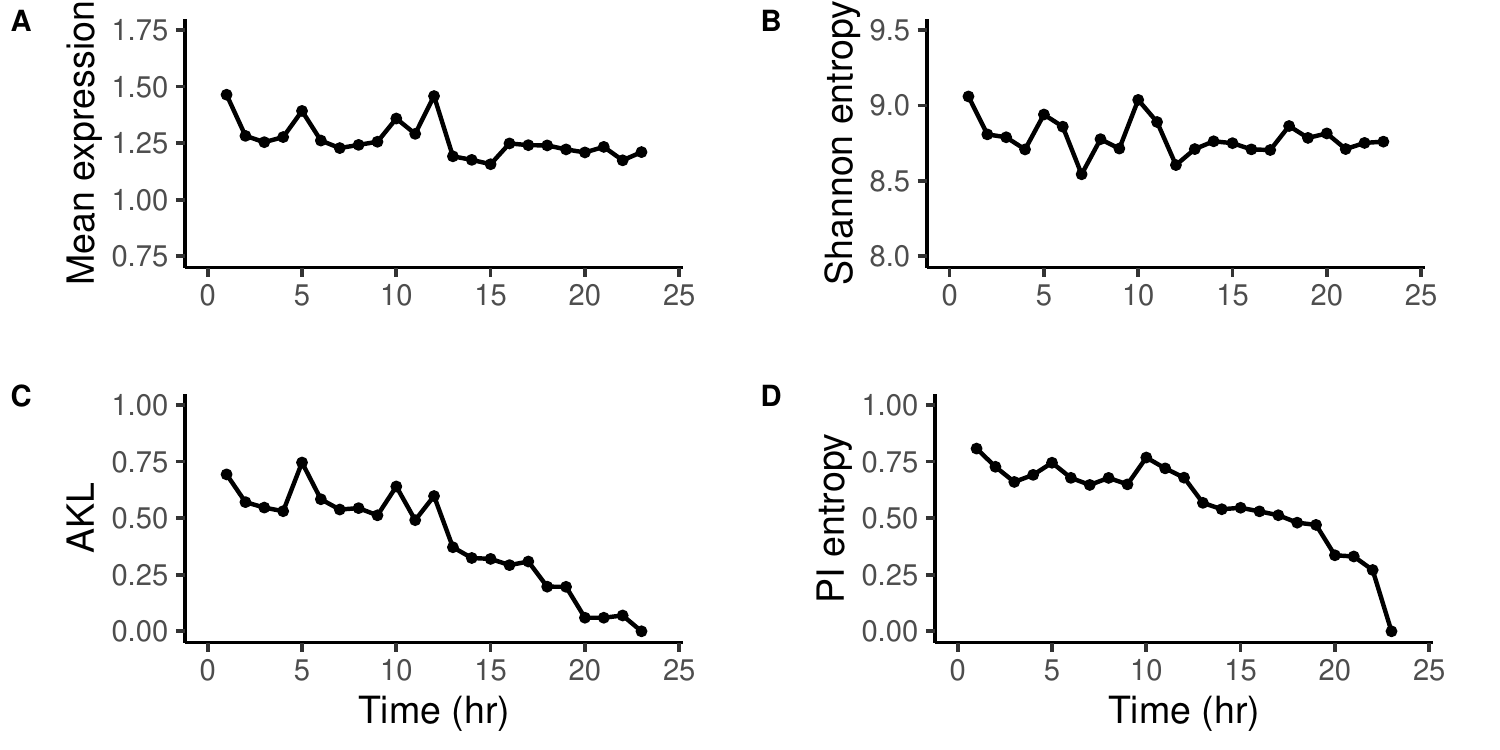}
    \caption{Statistical and information-theoretic properties of the gene ensemble expression as a function of time. (A) Mean expression value $\bar{\mu}$, (B) Shannon Entropy $H(t)$, (C) KL distance $A_{KL}$, and (D) PI entropy $\Pi(t)$. Mean expression is computed on relative gene expression values with the $\log_2$ transformation reversed to show greater dynamic range, but all genes are included in the analysis.}
    \label{fig:4PanelGrid}
\end{figure}

In Fig. \ref{fig:DistributionsByTime}, we see the time-development for the expression levels histograms (the "expression budget" or "energy" in our thermodynamic analogy), along with the locally smoothed value indicated by the black curve or 'envelope'.
While we can see some dynamical changes, they are not easy to describe precisely. However, the measures defined above allow us to capture global properties of these dynamics. In particular, we can see all the metrics as a function of time in Fig. \ref{fig:4PanelGrid}. Note in Fig. \ref{fig:4PanelGrid}A that the mean expression level $\bar{\mu}$ fluctuates with no obvious trend-line, except that it decreases overall, including in fluctuations, between the 10th and 15th hour.
Fig. \ref{fig:4PanelGrid}B shows the time-dependence of the Shannon entropy $H(t)$, which monitors the complexity of the envelope over the probability distribution $P(\mu(t))$. As with the mean expression level, here we also see fluctuations with no obvious trend-line, expect for an overall decrease, including in fluctuations, between the 10th and the 15th hour.

The other measures are more informative: the behavior of $A_{KL}(t)$ (Fig.~\ref{fig:4PanelGrid}C) shows an overall approach of the transcriptome to the final state $\vec{\mu(t_f)}$. The first time points generally have $\vec{\mu(t)}$ `far' from $\vec{\mu(t_f)}$. This difference decreases towards zero as we approach the final state, with a large final step.
Note that using $\vec{\mu(t_0)}$ as the reference point shows the opposite behavior (Fig.~\ref{fig:PI_with_t_0reference}): after an initial jump, gene expression is seen to generally remain a large distance from $\vec{\mu(t_0)}$ and is therefore not particularly informative. Since good reference points allow for maximum distinguishability between states, this further affirms the value of using $t_f$ for our reference ensemble here (see also below for $\Pi(t)$).

The $P(\mu)$ distribution envelope dynamics seen in Fig. \ref{fig:DistributionsByTime} 
are described via both its mean $\bar{\mu}(t)$ and complexity $H(t)$ metrics as going from initial greater fluctuations to settling at later times. To understand how this is consistent with the $A_{KL}$ dynamics seen, recall that these histograms do not capture the `secondary' dynamics of genes being shuffled or trading locations within an overall distribution that may not change much globally. Overall, these three metrics tell us that gene expression during development moves steadily in $\vec{\mu}$ space towards some sort of `settled' $\vec{\mu}(t_f)$, although it is unclear what exactly changes as the final state is approached. It is clear however that these changes happen masked behind the smaller changes happening to the overall expression envelope.

\begin{figure}[htbp]
    \centering
    \includegraphics[width=0.8\linewidth]{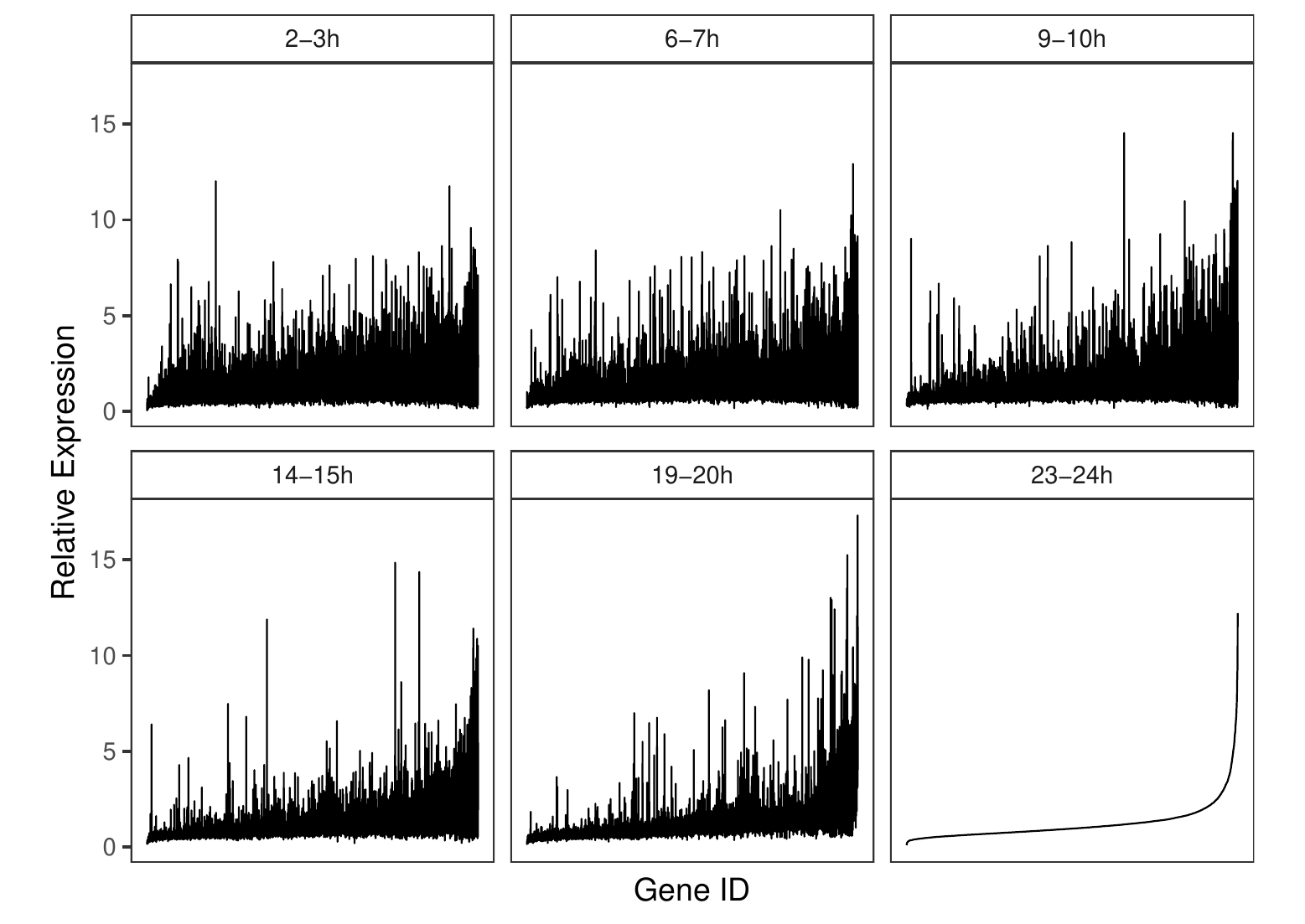}
    \caption{Relative expression values for each gene indexed by its final expression value. The panels here show the relative expression value for all genes as a function of time, where the gene is located along the horizontal axis (or indexed) according to its rank-ordered value as measured during the 23-24 hr window. We see the distribution settle down in overall range and relative smoothness, as quantified by $\Pi(t)$. Original expression values were reported with a $\log_2$ transform, here we have undone that.}
    \label{fig:Distributions}
\end{figure}

The $\Pi$ analysis proves extremely useful in unraveling what is happening during the `shuffling' in gene-expression level: In Fig.~(\ref{fig:Distributions}), we plot gene expression over time, where the $x$ axis is sorted according to $\mu(t_f)$. This makes $\tilde{\rho}(t_f)$ a reference smooth or `settled' distribution $\tilde{\rho}$ of the expression dynamics. This is quantified by $\Pi(t)$, shown in Fig.~\ref{fig:4PanelGrid}D, which exhibits an overall irreversible trend towards the final time $t_f$, like $A_{KL}$. Combining this behavior with all of the above, we can conclude that during development, expression levels of all these highly variable genes are shuffled around within an 
envelope of expression levels. The plots show that the expression distribution as well as its complexity settles to lower levels and lower fluctuations starting roughly at $t=15$~h, while the shuffling continues, proceeding in the direction of increasing smoothness or decreasing `braiding' in $\tilde{\rho}$, relative to the `final' state $\tilde{\rho} (t_f)$ at the end of development. 

\begin{figure}[h]
    \centering
    \includegraphics[width=0.8\linewidth]{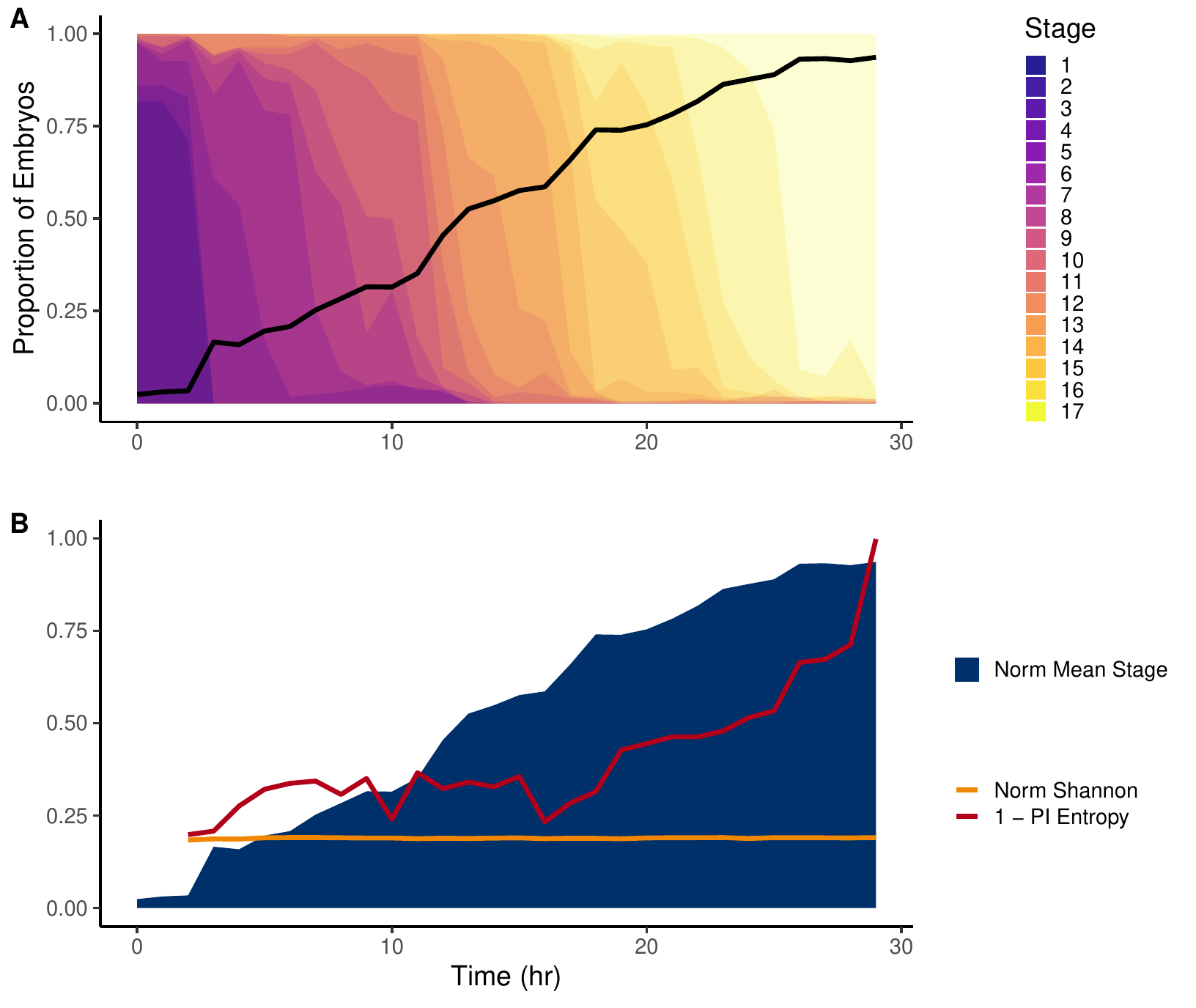}
    \caption{Relationship between time-development of ensemble properties and macroscopic or phenotypic properties. A) Proportion of sampled embryos in each of the 17 phenotypic stages.The black line indicates $M(t)$, the mean stage across all embryos at a given time, normalized to [0,1]. B) Mean stage (dark blue) compared to normalized Shannon entropy and PI entropy computed over all genes. PI is reported as 1-PI to match the stage progression. Note that in this figure time-points at $1/2$ hour marks are also being used. See text for details. }
    \label{fig:micro-Macro}
\end{figure}

\begin{figure}[h]
    \centering
    \includegraphics[width=0.75\linewidth]{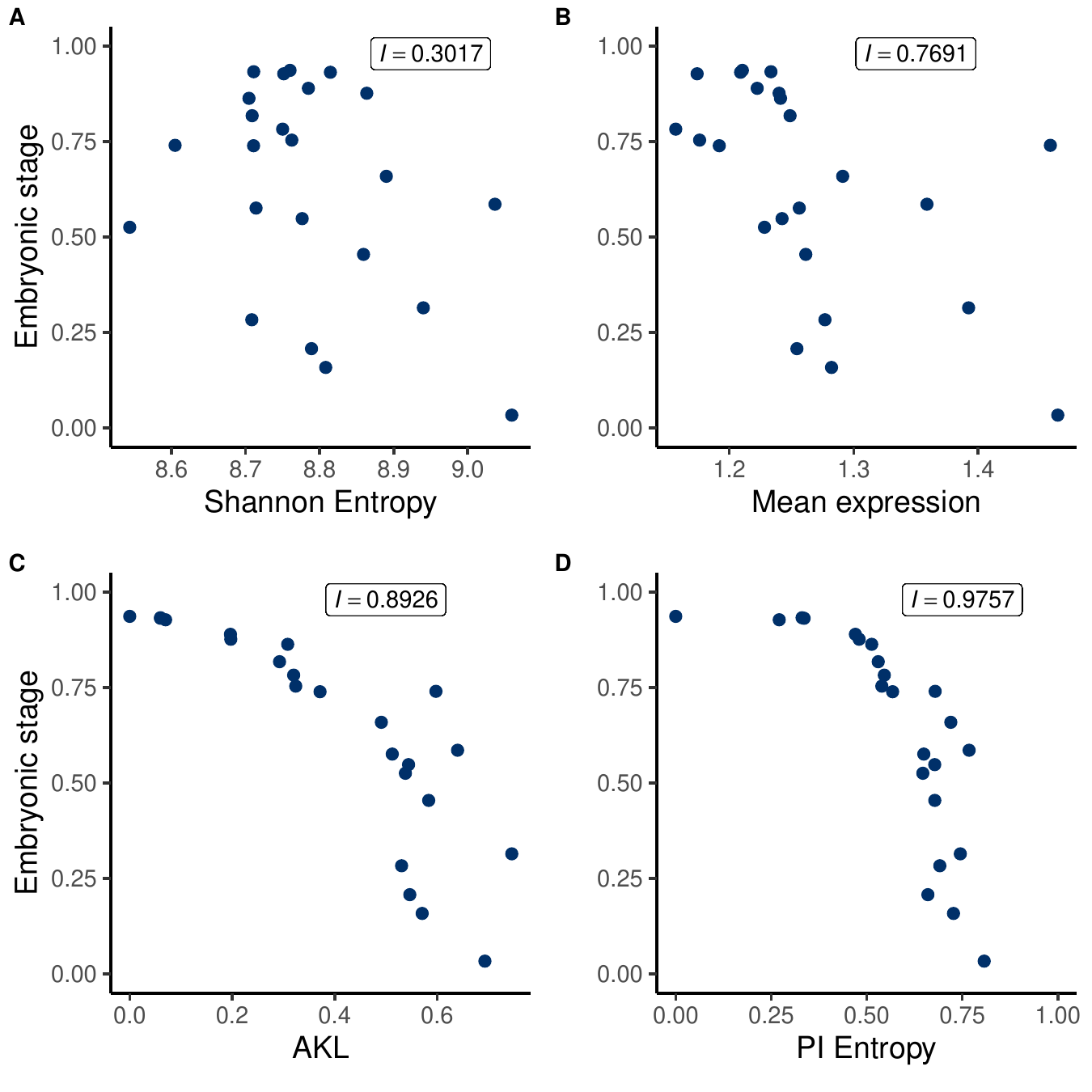}
    \caption{Relationships and mutual information between macroscopic phenotypic data and information-theoretic measures of gene expression data. These panels compare normalized mean embryonic stage $M(t)$ of the population against (A) Shannon entropy $H$, (B)  Mean expression level $\bar{\mu}$, (C) Approach Kullback-Leibler distance $A_{KL}$, and (D) PI entropy $\Pi$.} 
    \label{fig:mutinfo}
\end{figure}

Having thus plausibly demonstrated irreversibility as a property of gene expression dynamics, we turn to how this correlates with the developmental phenotypic state, a macroscopic property that is itself `irreversible' under typical biological conditions. Hooper et al. \cite{hooper_identification_2007} provide a distribution of qualitative phenotypic stages across the embryos harvested at each timepoint. Assuming that each developmental stage represents a different macrostate, we compute the mean phenotypic stage at each timepoint and use this as our macroscopic observable $M(t)$, referred to as embryonic stage and implicitly the normalized mean stage over the embryo population. Figure~\ref{fig:micro-Macro} shows both mean phenotypic larval stage $M(t)$ (Fig.~\ref{fig:micro-Macro}A) as well as $\Pi(t)$ and the Shannon entropy $H(t)$ (Fig.~\ref{fig:micro-Macro}B) 
as functions of time. Note that we actually graph $1-\Pi$ to clearly visualize the connection. We also note that for technical reasons related to the larval stage recording method, it is simpler and more consistent to use the complete dataset including the 'half time' points when doing this particular analysis (see Methods). This figure shows that there is indeed a strong correlation between the time-dependence of $M$ and the microscopic irreversibility measures above, and this correlation is quantified more concretely in Fig. \ref{fig:mutinfo}. Shannon entropy, perhaps the most common method for quantifying the complexity of a distribution, correlates poorly with embryonic stage, with a mutual information $I$ of $\sim$0.3 (Fig. \ref{fig:mutinfo}A). Mean expression achieves a higher mutual information ($I\approx 0.77$), but qualitatively still appears to correlate weakly with mean embryonic stage (Fig. \ref{fig:mutinfo}B). $A_{KL}$ and PI entropy correlate more strongly with embryonic stage both qualitatively and quantitatively (Fig. \ref{fig:mutinfo}C,D); PI entropy achieves the highest mutual information score of $I \approx 0.98$ (note that mutual information does not have a maximum of 1, this is not analogous to a Pearson correlation of 0.98). Both $A_{KL}$ and PI entropy broadly appear to have a negative exponential relationship with embryonic stage, but we refrained from fitting curves to these data due to the current lack of mechanistic understanding that would provide physical meaning to such an equation (see Discussion). This degree of correlation is remarkable considering the relatively coarse-grained nature of our transcriptomic data and the degree to which we are using proxies across multiple scales of coarse-graining. Using a finer-scale microscopic quantity, like single-cell RNAseq with embryo identity retained, we would expect even stronger correlations.

The correlation between the overall dynamics of the microscopic PI and the macroscopic mean phenotypic developmental stage shown in Fig. \ref{fig:mutinfo}C,D is our central result. 
This, as well as the form of the time evolution of both these quantities support our hypothesis that development is a time-directed process between non-equilibrium thermodynamic states, and that the corresponding thermodynamic trajectory is independently visible in both microscopic system state properties and macroscopic system properties.

A second way of understanding the results above is as validating the use of gene expression levels as a useful microscopic `coordinate' to characterize changes in biological state space. In particular, the behavior above suggests that $\mu$ can be thought of as a proxy for a concept such as energy, which is further reinforced by the connection between PI dynamics (indexed along the gene expression axis) and entropy. In other words, we see irreversible convergence of the ensemble dynamics of gene expression levels relative to a final indexed state
while these metrics also show broad distinguishability between gene expression states at various stages, as well as correlation with macroscopic dynamics. 

\subsection{Temporal and functional groups of genes, and particular development events}

\begin{figure}[ht]
    \centering
    \includegraphics[width=1\linewidth]{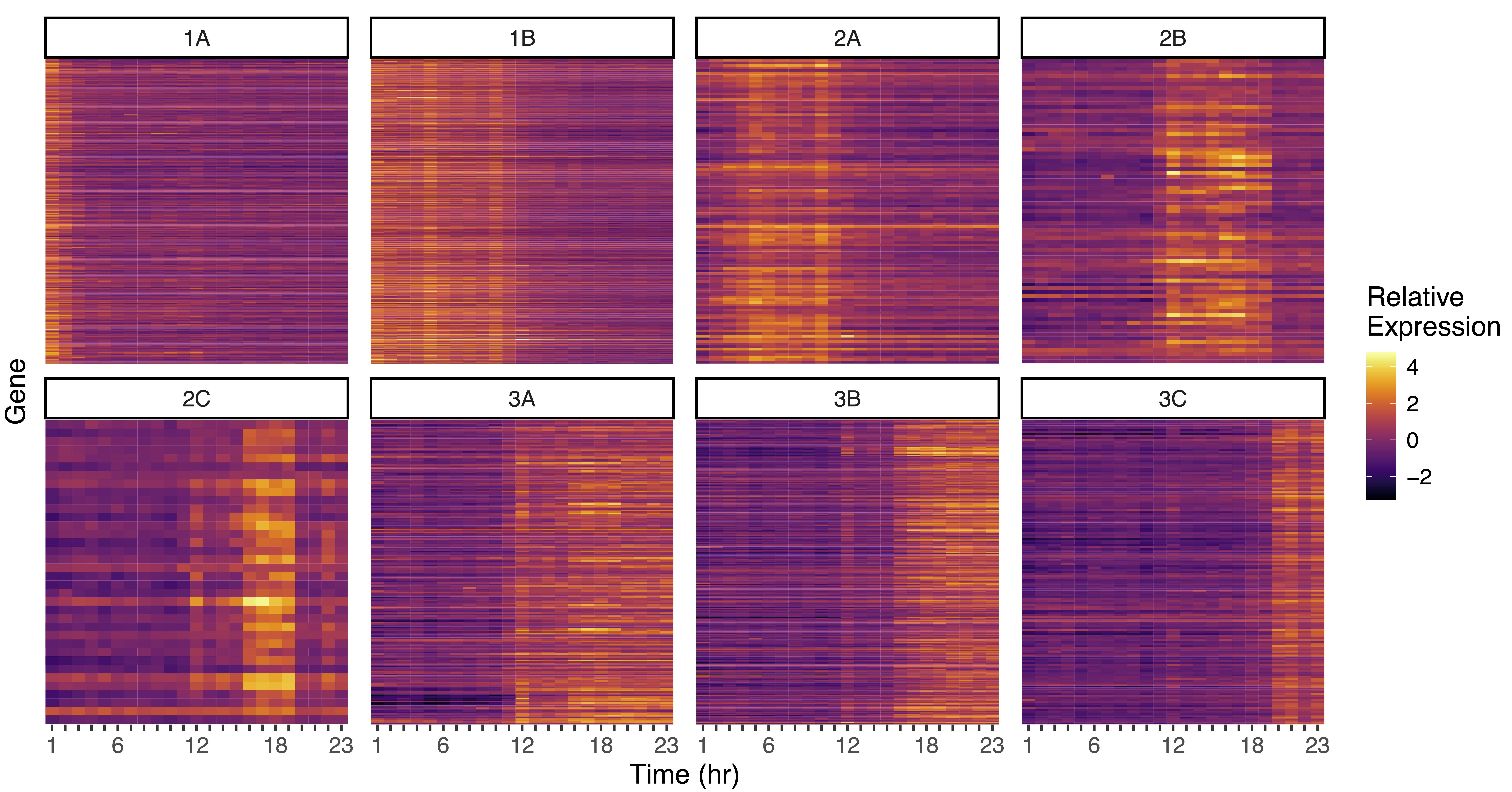}
    \caption{Expression level dynamics visualized as heat maps for the eight gene subgroups which display highly variable expression. These panels, reading across the two rows in sequence, show how certain genes are highly active only during certain time-periods, together making up the maternal (1A,1B), transient (2A,2B,2C), and activated (3A,3B,3C) classes, sourced from \cite{hooper_identification_2007}.}
    \label{fig:heatmaps}
\end{figure}

\begin{figure}[ht]
    \centering
    \includegraphics[width=0.8\linewidth]{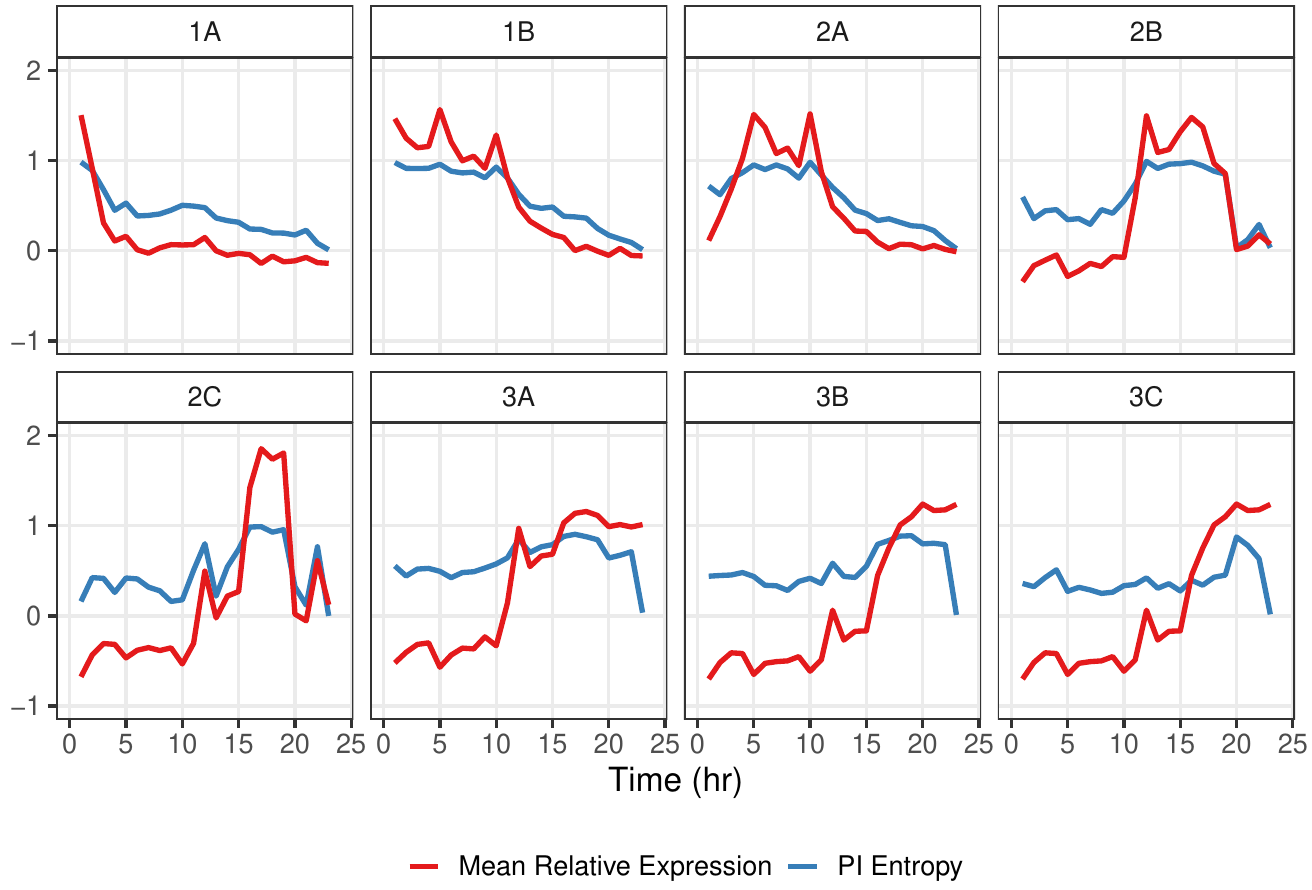}
    \caption{Mean relative expression and dynamics for the highly-variable subgroups. These panels show the computed mean relative expression $\bar{\mu(t)}$ for the eight gene subgroups shown in Fig.~\ref{fig:heatmaps} as a function of time over development. We can see that in general high expression is correlated with high $\Pi$ values, and that expression seems to 'travel' through different groups of genes. The relative expression is kept on the $\log_2$ scale to aid visual comparison.
    }
    \label{fig:pi_expression}
\end{figure}

The analysis presented in the previous section has been done without consideration of any biological details or microscopic features of the gene-expression dynamics. There are other ways of filtering the data to probe details of the thermodynamic trajectory, as well as its connection to macroscopic developmental stages. 
We first look separately at the dynamics of eight subgroups of genes with distinct time-resolved behavior, grouped into three gene classes that have been defined previously with separate biological interpretations \cite{hooper_identification_2007}:
\begin{itemize}
    \item \textbf{Class 1 (Maternal):} Genes with high initial expression that steadily decline over time. These genes are primarily maternally deposited transcripts that degrade as zygotic transcription takes over.
    \item \textbf{Class 2 (Transient):} Genes that exhibit a characteristic peak in expression during mid-embryogenesis before declining. These genes play roles in transient regulatory processes such as gastrulation and segmentation.
    \item \textbf{Class 3 (Activated):} Genes that are initially silent but increase in expression throughout development, typically encoding zygotic regulatory factors and structural proteins.
\end{itemize}
We examined the behavior of these gene sub-groups with high variability in time using heat-map visualizations of their expression dynamics, and by computing the same information-theoretic measures used for the global dynamics in the previous
section. The former analysis is shown in Fig.~\ref{fig:heatmaps}, and the latter in Fig.~\ref{fig:pi_expression}. 


In these figures we indeed see three different sorts of trajectories in gene expression space within the overall behavior described above: Genes from class I (Ia and Ib, the `maternal genes') start at high $\bar{\mu}$ and drop to lower $\bar\mu$ by the end. Class II genes (IIa, IIb, and IIc, the transient genes) start and end at low $\bar{\mu}$ with a spike or increase activity in $\bar{\mu}$ in between. Finally, Class III genes (IIIa, IIIb, and IIIc, the activated genes) start low and end high; recall that the overall $\bar\mu$ drops by this time. We observe expression progressing forward in time through the three classes, suggesting a model of these classes as belonging to temporally different `layers' of a feed-forward gene regulatory network \cite{gabalda-sagarra_recurrence-based_2018}. 

\begin{figure}[h]
    \centering
    \includegraphics[width=0.7\linewidth]{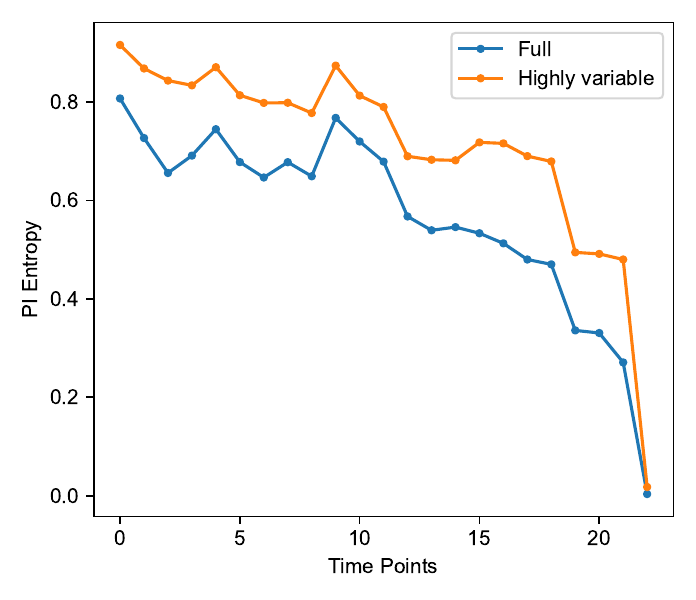}
    \caption{Comparison of dynamics for mean expression levels and $\Pi$ for the bulk relative to those for the highly-variable genes. These panels show $\Pi$ and mean values for the bulk whole transcriptome (complete data-set) as a function of time as well as for a merged group created from the $8$ subclasses showing high variability in expression levels, i.e from a single merged matrix which thus consists of `highly expressed genes' according to \cite{hooper_identification_2007}. We see that that bulk data and these highly-expressed subclasses show the same overall behavior.}
    \label{fig:bulkvhigh}
\end{figure}

In looking at the individual dynamics for these classes, we also see that increases in $\bar{\mu}$ are generally correlated with increases in $\Pi$, but these two measures are not saying the same thing. However, when we compare dynamics for $\Pi$ and $\bar{\mu}$ for the merged group of dynamically highly variable groups, we see---as in the bulk behavior---an overall decrease in $\bar{\mu}$, and a separate decrease in $\Pi$ complexity after the overall decrease. This is visible in Fig.~(\ref{fig:bulkvhigh}).
All of this is consistent with an overall analogy treating $\mu$ as an energy-like coordinate and $\Pi$ representing entropy-like properties consistent with that energy coordinate. In particular, all the expression dynamics - the global dynamics as well as the sub-class dynamics -- in general show that there is more `entropy' available at higher `energy' levels. However, we see two kinds of `relaxation' dynamics in gene expression space -- there is a `dissipation' down to lower `energy' $\bar{\mu}$ but there is also a separate `relaxation' time-scale visible in the the $\Pi$ dynamics, after the drop in $\bar{\mu}$. Thus, the overall irreversibility in the transition between the initial and final states of this developmental process can be thought of as proceeding first via a dissipation in expression level space along with a separate diffusive thermalization time-scale at that final level.

\begin{figure}[h]
    \centering
    \includegraphics[width=0.75\linewidth]{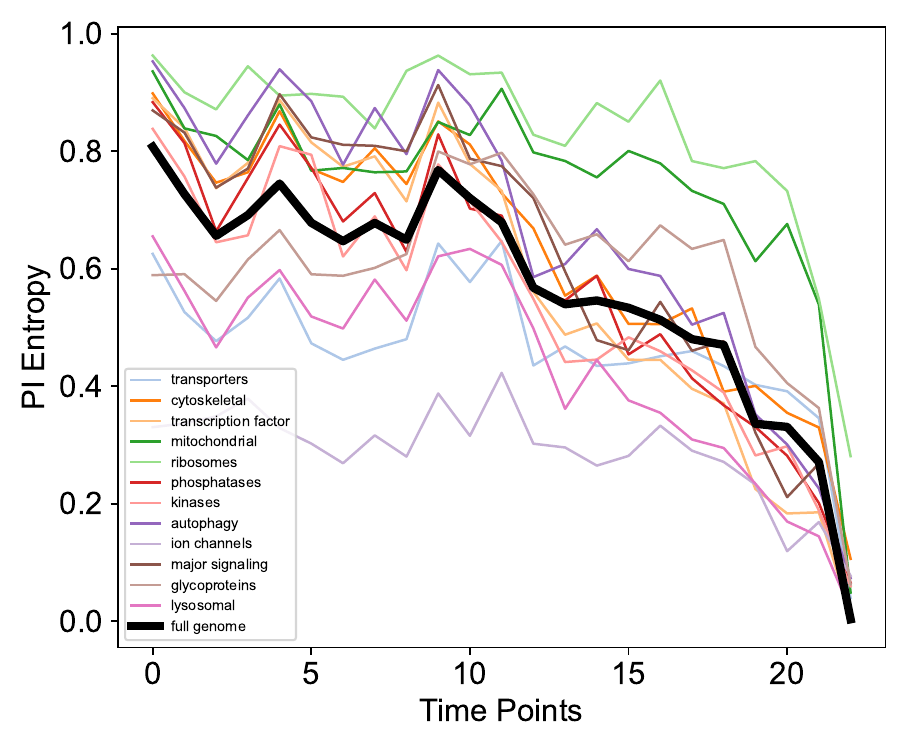}
    \caption{Applying $\Pi$ entropy analysis to gene expression dynamics for `functional' subclasses. Here we analyze the bulk whole-transcriptome RNAseq data but subdivided according to `functional' subclasses as labeled, each of which is analyzed separately. We see signs of a shared major `irreversible' event in the curves around the $9$th time point, close to the phenotypic symmetry-breaking time related to dorsal closure.}
    \label{fig:FunctionalClasses}
\end{figure}

We have also considered gene subsets created by filtering by functional (biological) gene classes sources from the GLAD database, as shown in Fig.~\ref{fig:FunctionalClasses}\cite{Hu2015GLAD}. It is clear that all the curves shown in Fig.~\ref{fig:FunctionalClasses} share certain common features.
We draw particular attention to overall increase in irreversibility signatures after about the $9$th time-point (visible as a spike in most gene groups), which is associated biologically with dorsal closure, arguably a critical point in cell-fate diversity and a symmetry-breaking event in the developmental process.

All gene classes roughly follow the dynamical trend of the whole transcriptome, although it is worth considering the two outlier gene groups, namely ribosomal genes, which maintain a high permutation entropy right until the final reference point, and ion channel related genes, which have a low entropy throughout development. It would tempting to conclude that the high PI levels of ribosomal genes reflect the embryo's need to maintain high translational flexibility as a downstream method of gene regulation, which has been shown to play an especially important role during development \cite{li_ribosome_2020, teixeira_translation_2021}. Similarly, the low PI of genes encoding ion channels could reflect the need for stable ion channel production rates due to the relevance of electrical signaling in development and pathologies associated with abnormal channel expression \cite{levin_bioelectric_2021}. Although using 28 noisy time points to characterize progression through 17 phenotypic larval stages may lead to over-interpretation, these observations can motivate new mechanistic studies at higher temporal resolution to further clarify the biological applicability of PI entropy.

\section{Discussion}
\label{sec:discussion}
This work serves as a test case for the broader statistical biophysics program of correlating dynamics of microscopic biomarkers, broadly construed, with macroscopic phenotypic behavior, and in particular for the hypothesis that such a correlation might be most easily found when thinking of development as a thermodynamic transition between non-equilibrium states. Our results suggest a transition along a \emph{sequence} of non-equilibrium states as the \textit{Drosophila} embryo develops. Further progress along this direction will require quantifying more carefully the specific functional form of the correlations between different scales, and in particular to work on understanding how these developmental trajectories vary at different scales of description, as discussed earlier. For example, it should be possible to go perform such analyses more precisely with finer resolution biomarkers such as single-cell RNAseq data, and similarly with better resolved phenotypic descriptions, such as examining cell-type diversity as a function of time. It would be generally useful to understand how the system's behavior or properties add or contribute at each scale. Further, useful next steps will include investigating the possible universality or individuality of these thermodynamic trajectories across species, taking into account as well variations in environmental conditions. The latter would be helpful in the considerable challenge of understanding how to quantitatively connect these properties to genuine thermodynamic properties we have labeled $E(t)$ above.

As such, we plan in future work to expand on these findings by applying the $A_{KL}$, PI entropy and other such information-theoretic tools to single-cell RNAseq data, which provides higher resolution and can distinguish cell-type-specific patterns, and to possibly connect these two levels of description. More ambitious plans will attempt to to integrate the PI entropy approach with specific gene-regulatory network models to quantify the influence of transcriptional regulators on developmental transitions. That is, by correlating the dependence of entropy measures with known gene regulatory networks, we aim to uncover causal relationships between gene expression fluctuations and morphological changes, ultimately seeking physically motivated equations relating these microscopic transcriptomic dynamics to macroscopic phenotypic ones, in the spirit of the connection between microscopic statistical mechanics and macroscopic thermodynamics (Fig. \ref{fig:mutinfo}).

We close by noting that beyond the demonstrated power of information-theoretic techniques applied to temporal RNAseq trajectories, this also serves as a test of the concept of PI entropy (Permutation entropy of an Indexed ensemble), recently introduced \cite{aragoneses_permutation_2023} as a framework for efficiently computing a proxy for the dynamics of correlations between members of an ensemble, relying on critical choices of indexing by an observable and using a reference point in time. In this context, it has proved to be useful in quantifying a `thermodynamic' trajectory that makes intuitive sense in the behavior of the dynamical complexity of gene expression during \textit{Drosophila} embryogenesis, as well as showing strong correlation with a proxy for the macroscopic phenotypic developmental stage. This highlights the utility of applying this sort of information-theoretic approach from non-equilibrium statistical mechanics or ensembles of dynamical systems to biomarker dynamics in developmental biology, with potential applications in other systems ranging from stem cell differentiation to evolutionary developmental biology.

\section*{Acknowledgments}
B.A. and M.G. thank internal Carleton College resources including the Eugster Student Research Endowment, the Towsley Endowment, and the Carleton Physics Department for supporting this research and its presentation, including travel. A.K.P. gratefully acknowledges internal sabbatical funding from Carleton College through a Presidential Fellowship and the Helms fund. A.K.P. also acknowledges the gracious hospitality of the Barcelona Collaboratorium for Modeling and Predictive Biology and the Dynamical Systems Biology group at the Universitat Pompeu Fabra for the opportunity to spend sabbatical time learning about these issues especially through very profitable talks and discussions with a wide variety of colleagues, in particular Alfonso Martinez-Arias. J.G.-O. was supported by the Ministerio de Ciencia, Innovaci\'on y Universidades and the Agencia Estatal de Innovación / FEDER (Spain) under project PID2024-160263NB-I00, and by the European Research Council, under Synergy grant 101167121 (CeLEARN). M.G. was also supported by the Fulbright U.S. Student Program during the preparation of this manuscript, which is sponsored by the U.S. Department of State and Spanish-American Commission. The contents of this manuscript are solely the responsibility of the author and do not necessarily represent the official views of the Fulbright Program, the Government of the United States, or the Spain Commission.

Additional thanks to Saayan Prasad for computational assistance. The authors have reviewed and edited the output and take full responsibility for the content of this publication.

\newpage
\section*{Supplementary Figures}

\renewcommand{\thefigure}{S\arabic{figure}}
\setcounter{figure}{0}

\begin{figure}[h]
    \centering
    \includegraphics[width=0.7\linewidth]{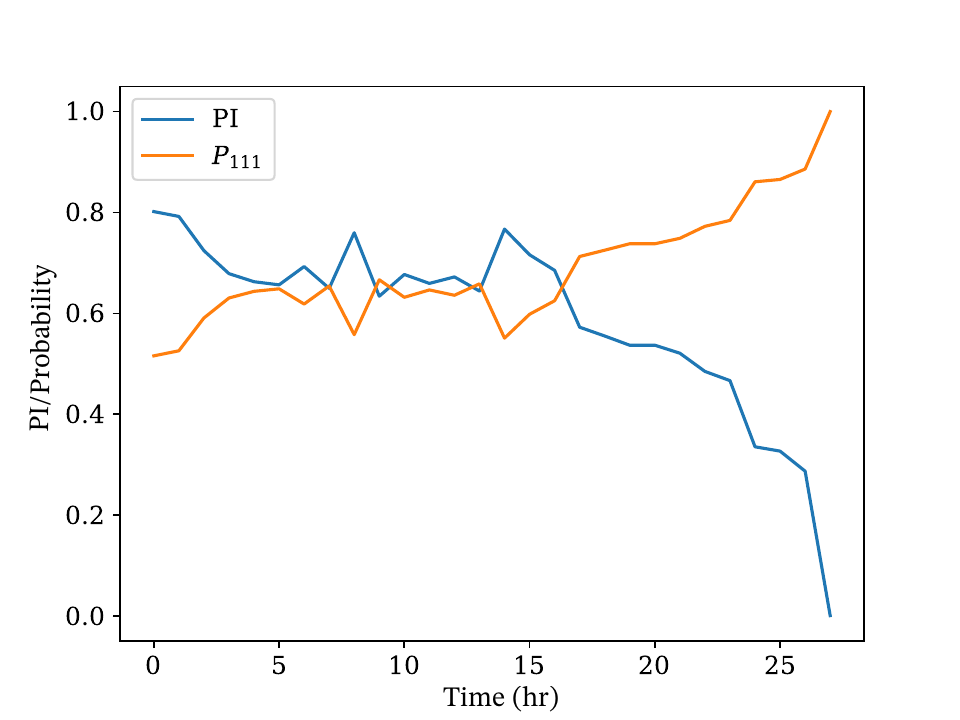}
    \caption{Relationship between $\Pi$ and $p_{111}$. We see here that for our situation, the growth in PI can be interpreted as arising from the growth of the $P_{111}$ population itself which happens due to the `tie' condition being applied to noisy data. See text for details.}
    \label{fig:PI-P111}
\end{figure}
\begin{figure}[h!]
    \centering
    \includegraphics[width=0.7\linewidth]{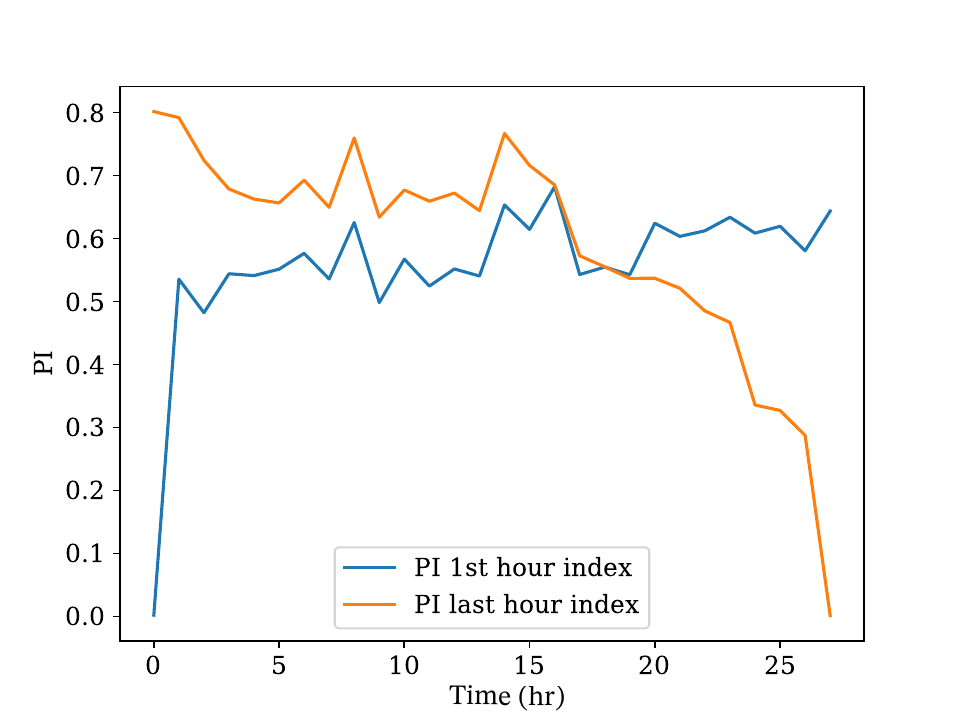}
    \caption{$\Pi$ indexed on first and last day}
    \label{fig:PI_with_t_0reference}
\end{figure}

\end{document}